\documentclass[letter]{jpsj2} 
\usepackage{bm}

\title{Spin Density Wave in Insulating Ferromagnetic Frustrated Chain LiCuVO$_4$}

\author{Takatsugu \textsc{Masuda}$^{1}$, Masato \textsc{Hagihala}$^{1}$, Yosuke \textsc{Kondoh}$^{2}$, 
Koji \textsc{Kaneko}$^{3}$, and Naoto \textsc{Metoki}$^{3}$}

\inst{$^{1}$Neutron Scattering Laboratory, Institute for Solid State Physics, 
the University of Tokyo, 106-1, Shirakata, Tokai, Naka-gun, 
Ibaraki, 319-1106, Japan \\
$^{2}$International Graduate Schools of Arts and Sciences 
Yokohama City University, Yokohama, 236-0027, Japan \\
$^{3}$Quantum Beam Science Directorate, Japan Atomic Energy Agency, Tokai, Ibaraki, 319-1195, Japan}

\abst{
We study field induced quantum phase in weakly-coupled ferromagnetic frustrated chain LiCuVO$_4$ by neutron diffraction technique. 
A new incommensurate magnetic peak is observed at $H \ge 8.5$ T. 
The field dependent propagation vector is identified with the spin density wave correlation in the theoretically predicted magnetic quadrupole order. 
Quantum fluctuation, geometrical frustration, and interchain interaction induce the exotic spin density wave long-range order in the insulating magnet.
}

\kword{ferromagnetic frustrated chain, nematic state, spin density wave}

\begin{document}
\maketitle
One of hot topics in condensed matter science is to search a spin liquid~\cite{Ramirez08} that exhibits order not in conventional two spin correlation but in others such as magnetic multipole~\cite{Andreev} or spin chirality.\cite{Kawamura} 
1D frustrated spin chain with ferromagnetic nearest neighbor interaction ($J_1$) and antiferromagnetic next nearest neighbor interaction ($J_2$) is a zoo of such novel states. 
While in case of the classical spin the spontaneous rotational symmetry breaking induces spiral long-range order (LRO) for $|J_1|/J_2~<~4$, 
those of quantum spin does not exhibit the LRO because of Mermin-Wagner theorem.\cite{Mermin66}
Instead solely the spin chirality, ${\bm \kappa} ={\bm S}_{i} \times {\bm S}_{j}$ does order with $Z_2$ symmetry broken and vector chiral (VC) phase appears.\cite{Kolezhuk05,Furukawa}

Applying magnetic field makes the system more complex and interesting. 
At the field close to the ferromagnetic polarized phase, a pair of magnons form bound state~\cite{Chubkov} and its Bose condensation at around $q = \pi$ induces quasi-LRO of the magnetic quadrupole.\cite{Kecke07} 
The quadrupole phase is characterized by the following longitudinal spin and transverse nematic correlations,\cite{Hikihara08} 
\begin{eqnarray}
\langle s_0^zs_l^z \rangle &\sim& M^2 - \frac{\eta}{\pi ^2 l^2} + \frac{A_z \cos (2k_F l)}{|l|^{\eta}} \\
\langle s_0^+s_1^+s_l^-s_{l+1}^- \rangle &\sim& \frac{A_m(-1)^l}{|l|^{1/\eta}}-\frac{\tilde{A_m}(-1)^l}{|l|^{\eta +1/\eta}}\cos (4\pi l) 
\label{correlation}
\end{eqnarray}
The former exhibits magnitude modulation with the characteristic wave vector $2k_{F}=2\pi \rho$, where 
$\rho$ is the density of bound two-magnon, $\rho=1/2(1/2-\langle S^z \rangle)$. $A_z$, $A_m$, and $\tilde{A_m}$ are positive constants and $\eta$ is Luttinger parameter.\cite{Hikihara08} 
Recent numerical studies exhibit magnetization vs $J_1/J_2$ phase diagram and the quadrupole phase in fact persists down to rather low magnetic field.\cite{Hikihara08,Sudan09,Meisner09} 
In addition the phase consists of two states, SDW2 in lower field where $\langle s_0^z s_l^z \rangle$ is dominant and nematic in higher field where $\langle s_0^+s_1^+s_l^-s_{l+1}^- \rangle$ is dominant. 
In both states transverse two spin correlation is short ranged and decays exponentially. 

In most quasi-1D magnet weak interchain interaction induces magnetic LRO at low temperature but it inherits quantum nature. 
In case of VC phase, spiral order in which the magnitude of the magnetic moment is strongly suppressed due to quantum fluctuation would be induced. 
In case of SDW2, LRO of the longitudinal spin correlation would appear with propagating wave vector $k_2 = 2k_F$.\cite{Sato09} 
The former is a good analogue for classical spin system but the latter is a totally novel state induced by frustration and quantum fluctuation. 

\begin{figure}
\begin{center}
\includegraphics[width=8.5cm]{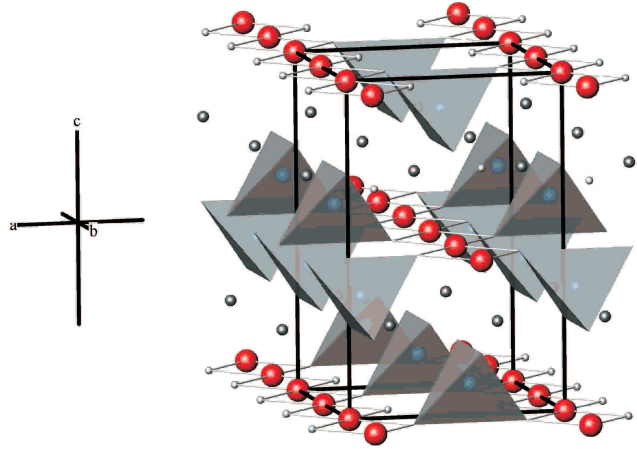}
\end{center}
\caption{(Color online) 
Crystal structure of LiCuVO$_4$. 
Cu-O chains separated by VO$_4$ tetrahedra and Li$^{+}$ ions are along the $b$ direction. 
$\angle$ Cu-O-Cu $\sim$ 90$^{\circ}$ indicates the ferromagnetic interaction. 
} \label{fig1}
\end{figure}

LiCuVO$_4$~\cite{Lafontaine} is one of the model compound for the frustrated ferromagnetic chain. 
As shown in Fig.~\ref{fig1} the CuO plaquette forms 1D $S$ = 1/2 chain in the crystallographic $b$ direction. 
Considering the bond angle of Cu-O-Cu $\sim 90^{\circ}$, nearest neighbor interaction is presumed to be ferromagnetic~\cite{Goodenough,Kanamori} and next nearest neighbor (NNN) interaction be antiferromagnetic(AF). 
The magnetic susceptibility showed typical behavior of 1D frustrated magnet, {\it i.e.,} broad maximum due to AF short-range fluctuation at $T_{\rm max}=28 {\rm K}$~\cite{Buttgen07} was observed. 
At $T \le T_{N}=2.3$ K incommensurate magnetic order with propagation vector ${\bm k_{\rm sp}}=(0~0.532~0)$ was identified.\cite{Gibson04} 
Neutron diffraction elucidates the spiral structure in the $ab$ plane at zero field~\cite{Gibson04} and also at small field $H \le 3.5$ T.\cite{Mourigal11} 
The magnetic moment is strongly suppressed as small as $0.25 \mu _{\rm B}$~\cite{Yasui08} $\sim 0.31 \mu _{\rm B}$.\cite{Gibson04} 
Inelastic neutron scattering showed enhanced spin dispersion in the $b^*$ direction and small one in others.\cite{Enderle07} 
Exchange parameters have been estimated from independent experiments including the magnetic dispersion,\cite{Enderle07} the continuum excitation,\cite{Enderle10} 
and magnetization measurements~\cite{Svistov11,Drechsler11} as summarized in the Table~\ref{parameters}. 
The realized spin model is weakly coupled ferromagnetic frustrated chains with NNN AF interaction as is presumed from crystal structure. 

\begin{table}
 \caption{Exchange parameters for LiCuVO$_4$. 
 $J_1$ and $J_2$ are nearest neighbor inchain and next nearest neighbor inchain interactions. $J'$ is interchain interaction. 
The difference between parameters in the Ref~\cite{Enderle10} and those in the Ref~\cite{Enderle07} is because the former was estimated from pure 1D chain model. 
 }
 \label{parameters} 
\begin{center}
 \begin{tabular}{l l l l l l}
& $J_{1}$~(meV) &  $J_{2}$~(meV) & $|J_{1}|/J_{2}$ & $J'$~(meV) \\
 \hline
INS (Ref.~\cite{Enderle07}) & -1.62 & 
3.56 & 0.29 & -0.408 \\
INS (Refs.~\cite{Enderle10}) & -2.4 & 3.4 & 0.71 &  \\
M-H (Refs.~\cite{Svistov11}) & -1.59 & 3.79 & 0.42 & -0.37 \\
M-H (Refs.~\cite{Drechsler11}) & -6.59 & 5.2 & 1.3 & \\
 \hline
 \end{tabular}
 \end{center}
 \end{table}

In magnetic field a phase transition was reported at a critical field $H_c \sim 7.5$ T~\cite{Banks07} and a spin-modulated collinear structure was proposed from the simulation of NMR spectra.\cite{Buttgen07,Buttgen10} 
The collinearity was confirmed also by the electric polarization measurement, {\it i.e.,} the ferroelectric polarization accompanied with the spiral order thorough spin current mechanism~\cite{Katsura05} disappears in the high field phase.\cite{Schrettle08} 
We think that this high field phase is related to SDW2 state in the magnetic quadrupole phase~\cite{Hikihara08,Sudan09,Meisner09} and the low field spiral phase is related to VC phase. 
For the statement, however, the direct observation of the spin correlation by neutron diffraction is required, which has been missed so far. 

In this paper we study the high field phase by neutron diffraction technique. 
With increasing field we found the disappearance of the $ab$ spiral structure and we observed the appearance of a new Bragg peak at $H \ge 8.5$ T. 
The propagation vector of the new peak has field dependence obeying $k= 2k_F$ and, thus, the new phase is identified with the SDW2-LRO induced by interchain interaction. 

For the sample preparation we used $^7$Li$_2$CO$_3$ isotope to reduce large neutron absorption by $^6$Li involved in natural Li$_2$CO$_3$. 
Single crystal of LiCuVO$_4$ was grown by traveling solvent floating zone method. 
LiVO$_3$ was used as a solvent and the growth speed was 0.1mm/h. 
The volume of the obtained crystal was 1.1cc. 
Magnetic susceptibility measurement was performed by commercial SQUID magnetometer. 
Preliminary neutron diffraction experiment was performed at PONTA spectrometer installed in JRR3 in JAEA and field experiment was performed at LTAS spectrometer in the same institute. 
Triple axis mode was used and the setup of sollar collimation was guide-80'-open-open. 
The neutron energy was 3meV and Be filter was used to reduce high energy neutron. 
Superconducting magnet was used to achieve 10T. 
The scattering plane was $a^*b^*$ plane. 

\begin{figure}
\begin{center}
\includegraphics[width=8cm]{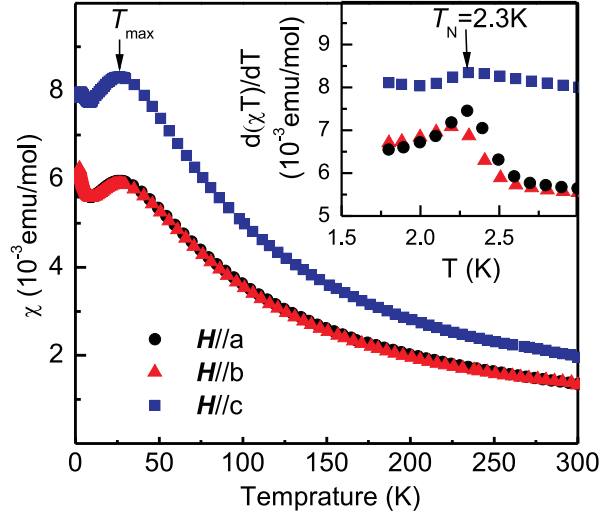}
\end{center}
\caption{(Color online) 
Magnetic susceptibility of $^7$LiCuVO$_4$ in field along $H \parallel$ the crystallographic $a$, $b$, and $c$ axis. 
The data for $H \parallel$ the $a$ and $b$ axes are overlapped. 
The inset is $d({\chi}T)/dT$ at low temperature region. 
} \label{fig2}
\end{figure}

Magnetic susceptibility $\chi$ of $^7$LiCuVO$_4$ is shown in Fig.~\ref{fig2}. 
AF fluctuation is enhanced at $T \lesssim T_{\rm max} \sim 26$K. 
Large magnitude of the $\chi$ in field applying to the $c$-axis is explained by anisotropy of g-tensor and easy-axis two-ion-anisotropy~\cite{Vasilev,Nidda}. 
At $T \le T_N$ the $\chi$ decreases in the $a$ and $b$ direction. 
N\'eel temperature is determined to be 2.3 K from the peak top of $d({\chi}T)/dT$,\cite{Fisher} which is consistent with previous report.\cite{Buttgen07} 

\begin{figure}
\begin{center}
\includegraphics[width=8cm]{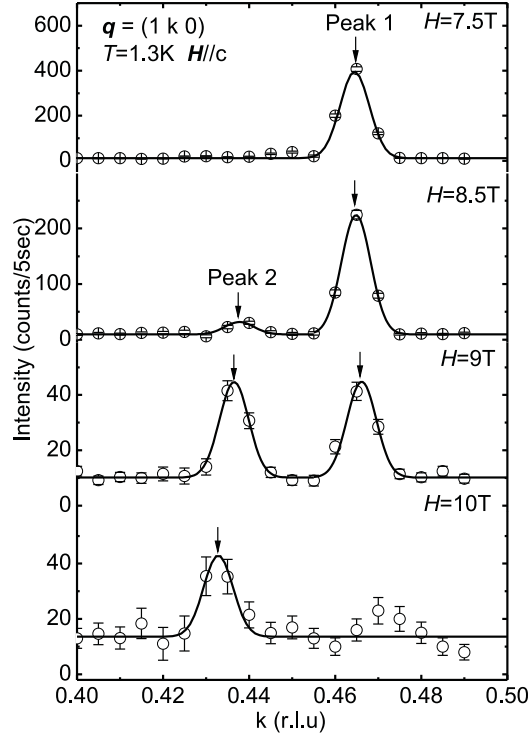}
\end{center}
\caption{
Neutron diffraction profile at ${\bm q}~=~(1~k~0)$ in $H \parallel c$ at $T$ = 1.3 K. 
The peak 1 is from $ab$ plane spiral structure and peak 2 is from spin density wave structure.  
} \label{fig3}
\end{figure}

To obtain the field dependence of magnetic Bragg peak neutron diffraction profiles at $q = (1~k~0)$ ($0\le k \le 1$) are collected at $T$ = 1.3 K. 
The field is applied perpendicular to the $ab$ spiral plane and there is no spiral plane flop as observed in $H~{\parallel}~a$ or $b$.\cite{Buttgen07} 
At $H$ = 7.5 T a peak (peak 1) is observed at $k \sim 0.465$ as shown in Fig.\ref{fig3}. 
The peak position is consistent with those obtained in previous experiments performed at $H \le 3.5$ T.\cite{Gibson04,Yasui08,Mourigal11} 
With the increasing $H$ the peak 1 is suppressed and a new peak (peak 2) appears at $k~\sim ~0.438$ at 8.5 T. 
Since the field agrees with phase boundary between spiral and collinear spin-modulated phases,\cite{Banks07,Buttgen07,Buttgen10,Svistov11} the peak 2 is ascribed to the collinear structure. 
With the increase of $H$ the intensity of the peak 2 increases and its position changes. 
Coexistence of two peaks at 8.5 T $\le H \le$ 9 T means that the phase transition is of 1st order. 
The peak 1 totally disappears at 10 T. 

\begin{figure}
\begin{center}
\includegraphics[width=8cm]{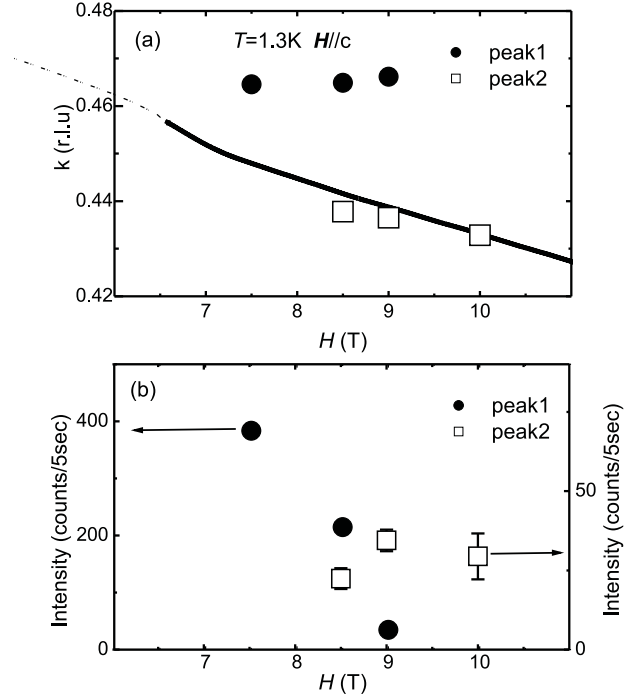}
\end{center}
\caption{(Color online) 
(a) The magnetic field dependence of Bragg peak position. Solid curve is $k_2 = (1-2\langle S^z \rangle )/2$ (see the text). The errorbars are inside the symbols. (b) The field dependence of Bragg peak intensities. 
} \label{fig4}
\end{figure}

The peak positions are plotted as a function of $H$ in Fig.~\ref{fig4}(a). 
The wave vector of the peak 1 is constant and the $ab$ spiral magnetic structure persists up to $H \sim 9$T. 
The wave vector of the peak 2 monotonically decreases with $H$. 
To discuss the quantum nature of the collinear phase, we draw the field dependence of the characteristic vector $k_2 = 2k_F$ of the SDW2 state in the magnetic quadrupole phase~\cite{Sato09} as solid curve. 
Consider that two Cu$^{2+}$ ions are along the chain direction in the unit length and it is obtained that ${\bm k_2}$ =  $(0~(1-2\langle S^z \rangle )/2~0)$. 
Here we use the magnetization data in Ref.~\cite{Svistov11} to estimate $\langle S^z \rangle$. 
The data is reasonably reproduced by the characteristic vector. 

The intensities for peak 1 and 2 are plotted in Fig.~\ref{fig4} (b). 
The intensity of the peak 2 is significantly suppressed. 
Here we will roughly estimate the magnitude of the magnetic moments.  
We assume SDW structure with ${\bm m_1} = m_{\rm SDW}(0~0~1)$ and ${\bm m_2} = m_{\rm SDW}(0~0~e^{-i\pi k_2})$ where ${\bm m_1}$ and ${\bm m_2}$ are the magnetic moment at the two Cu position in the chain and we simulate the Bragg intensity. 
Antiparallel spin alignment in the $[1/2~0~1/2]$ direction is required for the finite intensity. 
We also calculate the $I_{\rm spiral}(1~0.465~0)$ for the {\it ab} spiral structure, 
${\bm m_1}=m_{\rm sp}(1~i~0)$ and ${\bm m_2}=m_{\rm sp}(-e^{-i\pi k_{\rm sp}}~-ie^{-i\pi k_{\rm sp}}~0)$. 
Comparing the intensity of peak 1 at $H$ = 7.5 T and those of peak 2 at 10 T, it is found that $m_{\rm SDW}/m_{\rm sp}=0.29$. 

Field induced SDW-LRO was reported also in Ising-type quiasi-1D antiferromagnetic XXZ compound BaCo$_2$V$_2$O$_8$.\cite{Kimura08} 
In the isolated XXZ chain the zero field ground state is N\'eel state and applying field induces spin liquid with algebraic decay in spin correlation.\cite{Bogoliubov86} 
Unlike ferromagnetic frustrated chain the correlation is quasi-LRO both in longitudinal and transverse directions. 
The former is dominant in lower field region and, thus, in BaCo$_2$V$_2$O$_8$ SDW-LRO is induced in the mid-field region. 
The crucial difference is that there is no two-magnon bound state in the absence of ferromagnetic interaction. 
Instead one magnon is the dominant quasiparticle in low energy. 
Hence the density of one magnon, $\rho = (1/2-\langle S^z \rangle)$, should be used for the calculation of $\langle s_0^z s_l^z \rangle$ and then the characteristic vector is $2k_F = 2\pi (1/2-\langle S^z \rangle)$.\cite{Kimura08} 
In contrast in LiCuVO$_4$ it is the bound two magnon that dominates the low energy excitation. 
The density of two magnon is a half of one magnon, $\rho = 1/2(1/2-\langle S^z \rangle)$, and the propagation vector is differently expressed. 

In DMRG calculation~\cite{Hikihara08} the phase boundary between vector chiral and SDW2 phases are determined by the $M_c \equiv m(H_c)/m(H_{\rm sat})$ at which the magnetization step changes from $\Delta S^z=1$ to $\Delta S^z=2$. 
There is a discontinuous jump at $H_c$ for $\alpha \sim 2.4$ in the calculation size $L=168$~\cite{Hikihara08} and the authors predict the 1st order transition. 
Though the previous magnetization measurements did not detect hysteresis behavior~\cite{Banks07, Svistov11} probably because of the smallness, our neutron diffraction data clearly shows that the spiral and SDW phases coexists at $H_c$ and the phase transition is of 1st order. 

Previously reported value of $\alpha \equiv |J_1|/J_2$ is $0.3 \sim 1.3$ as summarized in Table~\ref{parameters}. 
In the theoretical $M$ - $\alpha$ phase diagram for ideal 1D chain, the phase boundary between vector chiral and SDW2 is $M_c \le 0.05$.\cite{Hikihara08,Sudan09} 
On the other hand from the magnetization measurement $M_c \sim 0.12$ was reported~\cite{Svistov11} and the calculated $M_c$ is underestimated. 
Furthermore easy axis anisotropy in LiCuVO$_4$~\cite{Nidda} stabilizes the SDW2 phase~\cite{Meisner09} and suppresses the $M_c$. 
For quantitative discussion on the phase diagram, theoretical calculation including interchain interaction would be necessary.  

In summary we study field induced quantum phase in a weakly coupled frustrated ferromagnetic chain compound LiCuVO$_4$ by neutron diffraction technique. 
New incommensurate Bragg peak is observed and its field dependence is explained by the characteristic vector of the spin density wave predicted in the magnetic quadrupole order in frustrated ferromagnetic chain. 
The phase is a consequence of quantum fluctuation, geometrical frustration, and interchain interaction. 

Recently long-range order of the nematic state is theoretically predicted close to $H_{\rm sat}$~\cite{Zhitomirsky10} and bulk magnetization shows an anomaly there.\cite{Svistov11} 
To study the spin dynamics in the high field nematic state by microscopic probe such as NMR would be an interesting future topic.

Dr. M. Sato and Prof. T. Hikihara are greatly appreciated for frutful discussion. 
This work was partly supported by Grant-in-Aid for Scientific Research (No.s
19740215 and 19052004) of Ministry of Education, Culture, Sports,
Science and Technology of Japan.


\end{document}